\documentstyle[epsf]{elsart} 
\begin{document}
\begin{frontmatter}
\title{Freezing of Simple Liquid Metals}
\author[Acadia]{A.R. Denton},
\author[Wien]{G. Kahl} and
\author[Wien]{J. Hafner}
\address[Acadia] {Department of Physics, Acadia University, Wolfville, NS, 
Canada, B0P 1X0}
\address[Wien] {Institut f\"ur Theoretische Physik and CMS,
TU Wien, Wiedner Hauptstra{\ss}e 8-10, A-1040 Wien, Austria} 
\date{Nov 20, 1998}

\begin{abstract}

Freezing of simple liquid metals and the relative stabilities of competing 
crystalline solids are investigated using thermodynamic perturbation theory, 
the interactions between ions being modeled by effective pair potentials 
derived from pseudopotential theory.
The ionic free energy of the solid phase is calculated, to first order in 
the perturbation potential, using classical density-functional theory and 
an accurate approximation to the hard-sphere radial distribution function.  
Free energy calculations for Na, Mg, and Al yield well-defined freezing 
transitions and structural free energy differences for {\it bcc, fcc,} and 
{\it hcp} crystals in qualitative agreement with experiment.

\end{abstract}
\end{frontmatter}

\section{Introduction} 

A microscopic description of freezing/melting phenomena presents a formidable
challenge to present-day condensed matter theory \cite{Low94-LH90}.
Among the simple liquids, liquid metals have received limited attention in 
this field, stemming from the relative complexity of their interatomic 
potentials and thermodynamic properties \cite{Han86}.
Previous related studies include those of Stroud and Ashcroft \cite{Str72}, 
who addressed the melting of simple metals using variational principles; 
Igl\'oi {\it et al.} \cite{Igl87}, who studied the freezing of liquid aluminum 
and magnesium by means of density-functional (DF) methods; and Moriarty and
coworkers \cite{Mor83}, who calculated the structural stabilities of 
third-period simple metals using first-principles techniques.

In this contribution, we report on a new study of the freezing phenomena
of simple metals that exploits recent advances in thermodynamic perturbation
theory for solids.  In line with Ref.~\cite{Igl87}, and most modern theories of
freezing, our approach incorporates classical DF theory \cite{Eva92} in
calculating the Helmholtz free energy.
Although formally rigorous, DF theory is limited in practice by insufficient
knowledge of the free energy functional for most except hard-body systems. 
To circumvent this limitation, we apply Weeks-Chandler-Andersen (WCA)
perturbation theory \cite{Han86,Wee71}, which splits the effective interatomic 
pair potential into reference and perturbative parts and maps the reference 
system onto an effective hard-sphere (HS) system.  For the inhomogeneous 
solid phase, DF theory provides an adequate approximation for the HS free 
energy, while an accurate model of the HS radial distribution function 
yields directly the first-order perturbation free energy. 
The homogeneous liquid phase is treated within the {\it same} perturbative
framework, using essentially exact expressions for the HS free energy and
pair distribution function.  This treatment of the liquid and solid by the 
same theoretical approach constitutes an important requirement for a consistent 
description of phase coexistence.

The next section outlines the theoretical methods used to calculate effective 
pair potentials and free energies of simple metals.  In Sec. 3 we present 
and discuss predictions of the theory for freezing transitions and relative 
stabilities of competing crystal structures.  
Finally, in Sec. 4 we summarize and conclude.

\section{Theoretical Methods}

Restricting attention to the simple metals, we construct effective 
density ($\rho$)-dependent interatomic 
pair potentials, $\phi(r; \rho)$, and a volume-dependent, but structure
insensitive, contribution to the binding energy, $E_V(\rho)$, 
via pseudpotential theory, assuming linear response 
of the conduction electrons to the ions.  In practice we use
empty-core pseudopotentials and the Ichimaru-Utsumi electron dielectric 
function, which are known to give reliable results for bulk liquid 
properties \cite{Haf87}.  
Taking $\phi(r; \rho)$ and $E_V(\rho)$ 
as the basic microscopic input, we proceed to calculate the 
thermodynamics of the liquid and solid phases by means of WCA perturbation 
theory.  Thus, we separate $\phi(r;\rho)$ at its first minimum into a 
short-range, purely repulsive, reference part, $\phi_{\rm o}(r)$, and a 
long-range,  weakly oscillating, perturbative part, $\phi_{\rm p}(r;\rho)$, and 
map the reference system onto an effective HS system with a HS diameter 
prescribed by the well-known WCA criterion.  This separation allows the 
ionic contribution to the Helmholtz free energy to be 
calculated within the coupling-constant formalism \cite{Han86}.  
To first order in $\phi_{\rm p}(r;\rho)$, the resulting general 
(inhomogeneous) expression for the total Helmholtz free energy reads

\begin{equation}
F[\rho({\bf r})] = F_{\rm HS}[\rho({\bf r})]
 + \frac{2\pi N^2}{V}\int_0^{\infty}{\rm d}r'~r'^2~
g_{\rm HS}(r';[\rho({\bf r})])~\phi_{\rm p}(r';\rho) + E_V(\rho),
\label{pert2}
\end{equation}

where $N$ is the number of particles, $V$ the volume, 
and $F_{\rm HS}[\rho({\bf r})]$ and $g_{\rm HS}(r;[\rho({\bf r})])$ 
the free energy and radial distribution function (RDF), respectively, 
of the HS reference system, both functionals of the equilibrium 
one-particle number density $\rho({\bf r})$.  The RDF of the solid is 
defined as an orientational and translational average of the 
two-particle density \cite{Ras96}.
For the liquid we use the homogeneous version of Eq.~(\ref{pert2}), taking
for the HS free energy and RDF the Carnahan-Starling and Verlet-Weis 
parametrizations of computer simulation data \cite{Han86}.

To calculate the free energy of the HS solid we invoke classical DF 
theory \cite{Eva92}.  The DF approach is based on the existence of a functional 
${\cal F}[\rho({\bf r})]$ of the density $\rho({\bf r})$ that satisfies
a variational principle, according to which ${\cal F}[\rho({\bf r})]$ 
is minimized -- for a given average density and external potential --
by the equilibrium density, its minimum value equaling the Helmholtz
free energy $F$.  In practice, ${\cal F}[\rho({\bf r})]$ is split into an 
(exactly-known) ideal-gas contribution ${\cal F}_{\rm id}[\rho({\bf r})]$ 
and an excess contribution ${\cal F}_{\rm ex}[\rho({\bf r})]$, 
the latter depending entirely upon internal interactions.  
Here we approximate ${\cal F}_{\rm ex}[\rho({\bf r})]$
by the modified weighted-density approximation (MWDA) \cite{Den89}. 
This approximation maps the excess free energy per particle of the solid 
onto that of a corresponding uniform fluid of effective density $\hat \rho$, 
according to 

\begin{equation}
{\cal F}^{\rm MWDA}_{\rm ex}[\rho({\bf r})] = N f_{\rm HS}(\hat \rho),
\label{MWDA}
\end{equation}

where the effective density, defined as

\begin{equation}
\hat \rho = \frac{1}{N} \int {\rm d} {\bf r} \int {\rm d} {\bf r'}
\rho({\bf r}) \rho({\bf r'}) w(|\bf r - \bf r'|; \hat \rho),
\label{rhohat}
\end{equation}

is a self-consistently determined weighted average of $\rho({\bf r})$.
The weight function, $w(r)$, is specified by normalization and by the
requirement that ${\cal F}^{\rm MWDA}_{\rm ex}[\rho({\bf r})]$ generates
the exact two-particle direct correlation function in the uniform
limit (see Ref.~\cite{Den89} for details). 

Practical calculation of ${\cal F}_{\rm HS}[\rho({\bf r})]$ and 
$g_{\rm HS}(r; [\rho({\bf r})])$ requires specifying the solid density,
{\it i.e.}, the coordinates of the lattice sites and the shape of the density
distribution about these sites.  Here we consider the {\it hcp}, {\it fcc}, 
and {\it bcc} crystals with the density distribution modeled by the usual 
Gaussian ansatz, introducing a parameter $\alpha$ determining the width 
of the distribution. This parametrization allows the ideal contribution to 
the free energy functional to be accurately approximated by 
${\cal F}_{\rm id}/N = (3/2)k_{\rm B}T\ln(\alpha\Lambda^2)-5/2$, 
where $\Lambda$ is the thermal de Broglie wavelength, and yields the HS 
free energy as the minimum with respect to $\alpha$ of the approximate
functional ${\cal F}_{\rm HS}[\rho({\bf r})]={\cal F}_{\rm id}[\rho({\bf r})]+
{\cal F}_{\rm ex}^{\rm MWDA}[\rho({\bf r})]$. 

For the RDF appearing in the perturbation contribution to the free energy 
[Eq. (\ref{pert2})] we adopt the approach of Rasc\'on {\it et al.} 
\cite {Ras96}.  Expressing $g_{\rm HS}(r;[\rho({\bf r})])$ as a sum over
coordination shells, this scheme approximates the distributions of the 
second and higher coordination shells in simple mean-field fashion, but
parametrizes the first-shell distribution in such a way as to 
incorporate nearest-neighbour correlations.  The first-peak parameters 
are determined by imposing sum rules, {\it e.g.}, the virial theorem and 
coordination number, so as to specify the contact ($r=d$) 
value and shape (area and first moment) of the first peak (for details 
see Ref.~\cite{Ras96}).  For a given solid structure, the lattice distances, 
the coordination number, the $\alpha$ that minimizes 
${\cal F}_{\rm HS}[\rho({\bf r})]$, and the HS pressure, derived directly 
from $F_{\rm HS}[\rho({\bf r})]$, then combine to determine 
$g_{\rm HS}(r;[\rho({\bf r})])$ and hence the perturbation free energy.

Following the above procedure, we obtain the {\it ionic} free energy, 
$F_{\rm ion}$, which is the only {\it structure}-dependent contribution
to the total free energy, and which suffices for assessing relative stabilities 
of different solid structures.  To fully describe freezing/melting, 
however, it is essential to consider the {\it total} free energy, 
$F=F_{\rm ion}+E_{\rm V}$, 
where $E_{\rm V}$ is an additional contribution from the 
conduction electrons.  The electronic free energy (or volume energy),  
$E_{\rm V}$, although independent of structure, depends on the average 
density and thus influences the densities of coexisting phases.  
Here we take for $E_{\rm V}$ the usual expression following from 
second-order perturbation theory \cite{Haf87} and consistent with 
the effective pair potential.

\section {Results and Discussion}

Figure 1 illustrates the density dependence of the effective HS diameter,  
$d$, for three different crystal structures of Al near its melting point.
According to the WCA prescription, $d$ depends both on the reference pair 
potential and on the first peak of the HS pair distribution function.  
Since the first peak is identical for {\it hcp} and {\it fcc} crystals, 
$d$ is the same for these two close-packed structures.  Differences in 
the first peak for the {\it bcc} crystal result in a smaller diameter for 
that more open structure.  Because of the sensitivity of $F_{\rm HS}$ to 
the HS packing fraction, these structural variations in $d$, though only 
of the order of 1 \%, can amount to significant variations in the free 
energy.  In passing, we note that the widely-used Barker-Henderson 
prescription for $d$ \cite{Han86} neglects such structural dependencies 
and that only knowledge of the HS RDF permits consistent application of 
the more accurate WCA prescription \cite{Ras96}.

Predicted structural free energy differences for Na, Mg, and Al 
are displayed in Fig. 2.  Evidently the theory correctly predicts the 
observed equilibrium structures for these three metals, namely {\it bcc}-Na, 
{\it hcp}-Mg, and {\it fcc}-Al, although the density at crossover from 
{\it fcc}- to {\it bcc}-Na is somewhat overestimated.  Compared with predictions
of first-principles approaches, our results agree qualitatively well.
For Mg and Al the stability trends of Fig. 2 match those from 
generalized pseudopotential theory (GPT) and the linear muffin-tin 
orbitals (LMTO) method \cite{Mor83}.  As well, for Na 
we predict the same trend as GPT, while LMTO fails for this case.
Compared with available experimental data, our structural free energy 
differences are in somewhat better quantitative agreement than those of 
either GPT or LMTO.  For example, for Mg and Al we predict differences
between {\it hcp} and {\it fcc} free energies of $-1.6$ and $3.2$ mRy, 
respectively, compared with $-0.6$ and $1.7$ mRy from GPT, $-0.7$ and 
$1.6$ mRy from LMTO, and $-1.5$ and $4.2$ mRy from experiment
(extrapolations of thermochemical alloy data, as quoted in Ref.~\cite{Mor83}).
It should be emphasized, however, that while the first-principles methods 
calculate ground-state ($T=0$) energies, thermodynamic perturbation theory
yields {\it free} energies of finite-temperature systems. 
We note further that application of perturbation theory is restricted 
to those structures for which the HS system is at least metastable, 
{\it i.e.}, for which ${\cal F}_{\rm HS}[\rho({\bf r})]$ has a local minimum.  
This excludes, {\it e.g.}, the diamond structure of Si, the HS diamond 
crystal being unstable.

Finally, we investigate the freezing/melting transition by comparing 
predicted free energies for the liquid and solid phases.  Figure 3 
exemplifies the comparison for Al near its observed melting temperature.
Crossing of the ionic free energy curves for the liquid and {\it fcc} 
phases implies a distinct transition near the observed equilibrium 
solid density.  Similar comparisons for Na and Mg also demonstrate 
liquid-solid phase transitions in qualitative agreement with observation.
In this respect, the effective-pair-potential approach reasonably describes
the many-body interacting system.
Note that the liquid-solid crossover density serves as a useful upper 
(lower) bound on the liquid (solid) density at coexistence.
Adding the volume-energy contribution -- identical for the two
phases -- while not affecting the crossover density, does alter the shape 
of the curves.  As a result, the densities of coexisting liquid and solid
phases, as determined by a Maxwell common tangent construction, depend
on $E_{\rm V}$.  An accurate volume energy is known to be essential 
for calculating realistic pressures and bulk moduli of metals 
from pseudopotential theory at the level of effective pair interactions 
\cite{Sin73,Fin74}.  Similarly, we find the liquid-solid coexistence analysis 
to be sensitive to the form of $E_{\rm V}$, particularly its curvature 
with respect to $\rho$.  In fact, the sensitivity is such as to preclude 
here a reliable determination of coexistence densities.  Given the reasonable
predictions for ionic free energies, this apparent limitation of the theory 
may reflect a need to go beyond the linear response approximation to include 
higher-order response effects in the volume energy (see also
\cite{Haf87}, p. 42ff). 
The relative significance of such effects, as well as distinctions~\cite{Fin74}
between the real-space representation of $E_{\rm V}$ used here and the 
reciprocal-space representation of Ref.~\cite{Str72}, are interesting issues 
deserving of further attention.

\section{Conclusions}

Summarizing, working within an effective-pair-potential framework, we have 
implemented a consistent form of thermodynamic perturbation theory, 
one that is of comparable accuracy for both the liquid 
{\it and} solid phases, and used it to study structural stabilities and 
freezing behaviour of third-period simple metals.  Predictions for structural 
free energy differences are at least as accurate as those from two 
first-principles methods and are in generally good
qualitative agreement with experiment.  On the basis of ionic free energies, 
freezing/melting transitions are predicted near the observed densities.
As the densities of coexisting liquid and solid phases are found to depend
sensitively on the electronic free energy, a complete coexistence analysis 
has not been attempted pending more accurate knowledge of this part of 
the free energy.  The computationally practical approach 
demonstrated here should prove useful in future studies of the freezing 
of liquid metal mixtures, as well as in guiding more sophisticated 
{\it ab initio} simulations. 

{\bf Acknowledgements}

ARD acknowledges the Forschungszentrum J\"ulich for use of its computing
facilities.  GK and JH acknowledge the support of the \"Osterreichische 
Forschungsfonds under Proj. Nos. P11194 and P13062 and of the 
Oesterreichische Nationalbank under Proj. No. 6241.

\newpage

\begin{figure}
\caption{WCA effective hard-sphere diameter in atomic units (Bohr radii) 
for Al at temperature $T=930$ K in {\it hcp, fcc}, and {\it bcc} crystal 
structures.}
\label{fig1}
\end{figure}

\begin{figure}
\caption{Free energy differences between various crystal structures vs.
reduced atomic volume, where $\Omega_{\rm o}$ is the observed equilibrium 
atomic volume at atmospheric pressure: (a) Na at $T= 300$ K; 
(b) Al and Mg at $T= 930$ K.}
\label{fig2}
\end{figure}

\begin{figure}
\caption{Ionic free energy vs. average density for liquid and 
{\it fcc}-solid phases of Al at $T= 930$ K.}
\label{fig3}
\end{figure}

\end {document}